\documentclass[prb,floatfix,twocolumn,letterpaper,nobalancelastpage]{revtex4-2}
\usepackage{mycommands}

\begin{document}
    
    \title{Terahertz Circular Dichroism in Commensurate Twisted Bilayer Graphene}
    
    \author{Spenser Talkington}
    \affiliation{Department of Physics and Astronomy, University of Pennsylvania, Philadelphia, Pennsylvania 19104, USA}
    \email{spenser@upenn.edu}
    
    \author{Eugene J. Mele}
    \affiliation{Department of Physics and Astronomy, University of Pennsylvania, Philadelphia, Pennsylvania 19104, USA}
    
    \date{\today}
    \begin{abstract}
        We report calculations of terahertz ellipticities in large-angle, 21.79$^\circ$ and 38.21$^\circ$, commensurate twisted bilayer graphene, and predict values as high as 1.5 millidegrees in the terahertz region for this non-magnetic material. This terahertz circular dichroism is surprisingly large and has a magnitude comparable to that of chiral materials in the visible region. At low frequencies, the dichroic response arises from the strong interlayer hybridization, which allows us to probe the symmetry and strength of these couplings. Crucially, lateral interlayer translation tunes this response, in striking contrast to the near invariance of small twist angle bilayer to interlayer translation. We examine the magnitude and phase of the interlayer coupling for all structures containing fewer than 400 atoms per unit cell. Finally, we find that the dichroism can be manipulated by applying an electric field or with doping.
    \end{abstract}
    
    \maketitle

    \section{Introduction}\label{section:intro}

        Twisted bilayer graphene is a chiral material whose chirality originates with the direction of the interlayer twist. This chirality manifests in optical measurements as a twist-dependent circular dichroism \cite{kim2016chiral}. This circular dichroism is the natural optical activity of a time-reversal symmetric system and scales linearly with frequency as $\Omega d/c$ for thickness $d$ and speed of light $c$ \cite{landau2013electrodynamics}. A priori, this linear scaling should strongly suppress circular dichroism in the terahertz spectrum. We show that while such a supression does indeed occur when the electronic states are approximated as Dirac fermions, this description breaks down at low energy scales in commensurate twisted bilayer graphene (C-TBG). In C-TBG, interlayer scattering processes hybridize electronic states \cite{shallcross2008quantum,mele2010commensuration,mele2012interlayer} and as a consequence a robust THz dichroism occurs in these materials. In essence, the strong interlayer hybridization preempts the Dirac approximation at low frequencies and thereby avoids the suppression from small photon wavevectors. At optical frequencies the effect of interlayer hybridization is smaller and the problem reverts to the usual description of circular dichroism in a Dirac band which has been studied in previous work.
        
        Quantitatively, we predict ellipticities as high as 1.5 mdeg for commensurate twisted bilayer graphene in the terahertz region. This is comparable to the peak of 4.3 mdeg observed in incommensurate graphene bilayers \cite{kim2016chiral} and 3.1 mdeg predicted in commensurate bilayers but measured at visible frequencies \cite{addison2019twist}, in contrast to the suppression with $\Omega$ expected for natural optical activity. While there is a suppression with $\Omega$, at low energies, current operator matrix elements become large and counteract the suppression with $\Omega$. Adjusted for thickness, this terahertz circular dichroism is 26 times larger than a similar thickness of D-glucose in the visible region \cite{matsuo2004vacuum}.

        Twisted bilayer graphene is a tunable platform for realizing exotic low-energy physics. Five years ago, superconductivity was observed in small twist angle bilayer graphene \cite{cao2018unconventional,yankowitz2019tuning}, and since then many other correlated electronic phenomena have been observed \cite{andrei2020graphene}. Since then studies have looked for and found twist-induced electronic properties in transition metal dichalcogenide bilayers and other moir\'e materials \cite{andrei2021marvels}. Until recently the primary focus in moir\'e materials has been on long repeat period and quasiperiodic systems that naturally realize long wavelength physics such as the flat bands \cite{trambly2010localization,bistritzer2011moire} where strong-correlation physics occurs. These moir\'e structures stand in sharp contrast to short repeat period commensurate structures. In C-TBG interesting non-Dirac physics can occur such as flat bands \cite{kindermann2012zero,pal2019emergent,scheer2022magic}, or as we study here interlayer hybridization on the terahertz energy scale.
        
        Recently large twist angle bilayer graphene has reentered the spotlight with the observation of superconductivity
        \cite{zhou2022isospin}, and the quantum anomalous Hall effect \cite{geisenhof2021quantum} in Bernal bilayers which are the {\it shortest} period stacked structures. C-TBG, whose twist angle leads to a periodic real space structure, can be viewed as inflated versions of AA and Bernal stacked bilayer graphene with a reduced energy scale \cite{shallcross2008quantum,mele2010commensuration,mele2012interlayer,talkington2023electric}. 
        At incommensurate angles the large angle twisted bilayer graphene's electronic structure can be modeled with independent Dirac cones residing in the two layers \cite{dos2007graphene,neto2009electronic,rozhkov2016electronic}, this description breaks down in commensurate structures where reciprocal lattice vectors connect the valleys of the two layers and hybridize the low energy structure \cite{mele2010commensuration,weckbecker2016low}. This hybridization energy scale is crucial in determining the energy window over which universal Dirac physics is violated. In the following we obtain the scattering amplitude and phase for all C-TBG structures with less than 400 atoms in the unit cell.

        While the responses of TBG to linearly \cite{morell2012radiation,tabert2013optical,moon2013optical,stauber2013optical,le2018electronic} and circularly polarized light \cite{morell2012radiation,kim2016chiral,morell2017twisting,stauber2018chiral,addison2019twist,do2020optical} have received substantial attention, the terahertz response properties of TBG have been relatively unappreciated. Ten years ago Ref. \cite{zou2013terahertz} studied the terahertz optical response of bilayer graphene, but the samples were made by chemical vapor deposition and one wavelength of light covered regions with many twist angles. Since then material quality and tunability has substantially improved, so probing unusual optical physics in these samples in the terahertz region has become more practical.
        Experiments on the terahertz magneto-optical activity of a monolayer \cite{nedoliuk2019colossal}, and terahertz circular dichroism measurements in gated Bernal bilayers have subsequently been conducted \cite{zhang2021actively,wang2022independently}, but these studies do not take advantage of the twist degree of freedom. Recently, Ma \textit{et al} studied the infrared optical properties of gated, twisted double bilayer graphene in Ref.  \cite{ma2022intelligent}.
        We showed in Ref. \cite{talkington2023electric} that C-TBG exhibits a field-tunable band gap and has a power-law divergence in the optical conductivity at the band edge, and Ref. \cite{jiang2022tunable} showed that the band gap in C-TBG is tunable with strain. The results on twisted bilayer graphene might be extended to doped semiconducting materials such as twisted hexagonal boron nitride \cite{ochoa2020flat}.
        Here we show that the tunability of C-TBG can be used to selectively realize a large circular dichroism in the terahertz region in the absence of magnetism.

        In Section \ref{sec:model}, we present commensuration conditions, use fit a tight-binding model to obtain interlayer scattering amplitudes and phases, present a symmetry based model for the low energy electronic degrees of freedom, and emphasize its salient features.
        In Section \ref{sec:response}, we present the response function and derive current operators as they apply to this problem.
        In Section \ref{sec:cd}, we present our main results on the optical conductivity and circular dichroism (ellipticity) in C-TBG as a function of photon frequency and perpendicular electric field.
    
    \section{Model}\label{sec:model}

        At a dense set of interlayer twist angles, twisted bilayer graphene exhibits commensurate unit cells \cite{campanera2007density}, however most of these unit cells have long repeat periods and the interlayer coherence energy scale that dictates the low energy band structure. This scale falls off rapidly with increasing unit cell size \cite{shallcross2008quantum}. Here we consider all commensurate structures with less than 400 atoms in the unit cell and use a tight-binding model to determine the interlayer coupling scale. We conclude that only the $21.79^\circ$ and $38.21^\circ$ structures will
        exhibit interlayer coherence scales in the terahertz region, with strong power law divergences in the optical conductivity \cite{talkington2023electric} and show measurable circular dichroic behavior in this energy window. We then present a symmetry-based model for the low energy electronic behavior of C-TBG. For the current operators and optical response, proceed to Sections \ref{sec:response} and \ref{sec:cd} respectively.

        \subsection{Commensuration Conditions}

            \begin{figure}
                \centering
                \includegraphics[width=\linewidth]{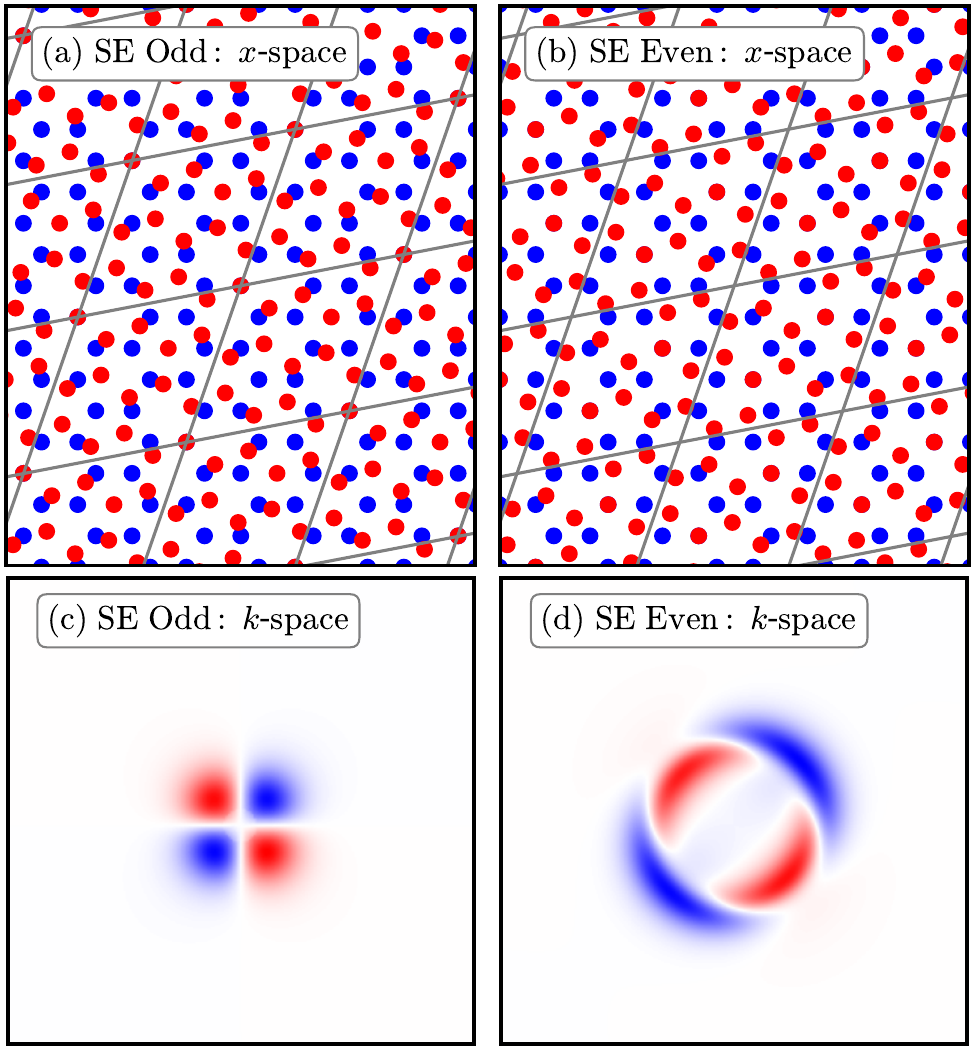}
                \caption{SE-odd and SE-even commensurate twisted bilayer graphene (C-TBG) differ by a small interlayer translation; visualized for the $\sqrt{7}\times\sqrt{7}$ structures. \textbf{(a)} The SE-odd structure exhibits points with at most $C_3$ symmetry. \textbf{(b)} The SE-even structure exhibits points with $C_6$ symmetry. \textbf{(c)} Momentum space contributions to $\text{Im}(\sigma^{xy})$ reveal the $l=2\hbar$ phase winding of the wavefunctions about the $K$ point in the SE-odd structure. \textbf{(d)} Interlayer shift makes the wavefunctions a superposition of angular momentum eigenstates and circular dichroism appears as visualized for the SE-even structure.}\label{FIG:real28}
            \end{figure}

            \begin{figure*}
                \centering
                \includegraphics[width=\linewidth]{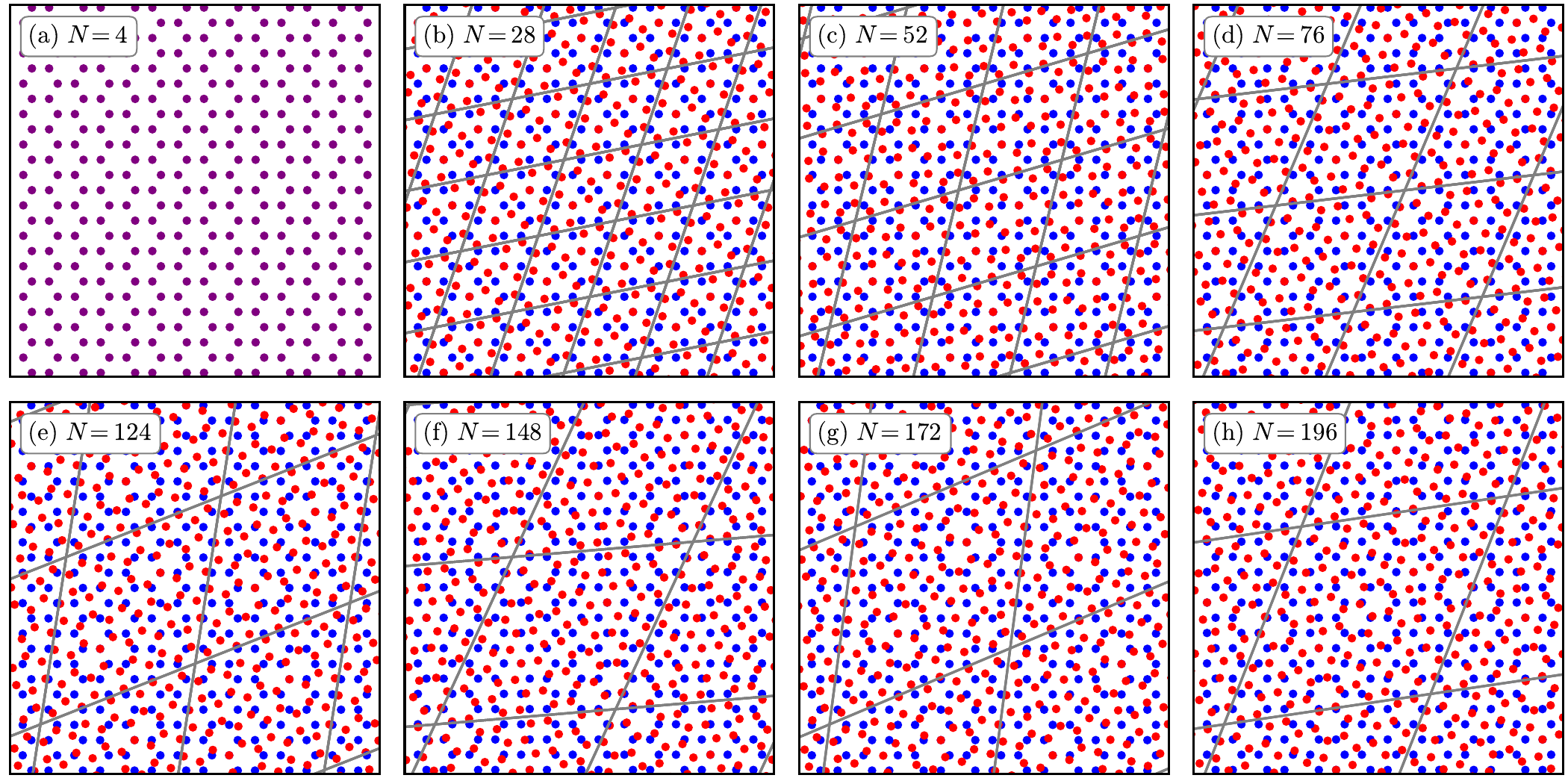}
                \caption{Real space superlattice structures of the eight smallest SE-even C-TBG systems. The system with $N=4$ atoms is the AA stacked bilayer. The gray lines outline the unit cells and intersect at points of $C_6$ symmetry. The existence of these points is key to the sublattice symmetry; for a generic twist and shift of graphene bilayers there will be no points with $C_6$ symmetry.}\label{FIG:realeven}
            \end{figure*}

            In monolayer graphene we have primitive translation vectors $t_1^{(1)}=a e^{-i\pi/6}$ and $t_2^{(1)}=a e^{i\pi/6}$; for a second graphene monolayer with twist angle $\theta$ relative to the first layer we have translation vectors $t_1^{(2)}=a e^{i\theta}e^{-i\pi/6}$ and $t_2^{(2)}=a e^{i\theta}e^{i\pi/6}$. The commensuration condition is
            \begin{align}
            m^{(1)} t_1^{(1)} + n^{(1)} t_2^{(1)} = m^{(2)} t_1^{(2)} + n^{(2)} t_2^{(2)}.
            \end{align}
            Now, for commensuration to occur, it must be the case that $m^{(1)}=n^{(2)}:=m$ and $n^{(1)}=m^{(2)}:=n$ \cite{mele2010commensuration,mele2012interlayer}, so the commensuration condition becomes
            \begin{align}
            m t_1^{(1)} + n t_2^{(1)} = n e^{i\theta}t_1^{(1)} + m e^{i\theta}t_2^{(1)},
            \end{align}
            which means that the commensurate angles are given by
            \begin{align}
            \theta = \text{Arg}\left(\frac{m e^{-i\pi/6} + n e^{i\pi/6}}{n e^{-i\pi/6} + m e^{i\pi/6}}\right),
            \end{align}
            which are indexed by $m$ and $n$.
        
            The structures are classified by their sublattice exchange (SE) symmetries. SE-even structures have points with $C_6$ symmetry while SE-odd structures have points with at most $C_3$ symmetry. For examples of SE-even and SE-odd real-space structures with 28 atoms per unit cell, see Fig. \ref{FIG:real28} (a-b). SE-even structures are those for which
            \begin{align}
            \frac{n-m}{\text{gcd}(m,n)}\ \text{mod}\ 3 = 0,
            \end{align}
            and are rotated about a point where an atom in the $A$ sublattice of the second layer lies above an atom of the $A$ sublattice in the first layer. The sublattice even structures also have places where atoms on the $B$ sublattices of both layers overlap \cite{mele2012interlayer}: call this location $\bm{r}_{BB}$, then $2\bm{r}_{BB}$ is a point with $C_6$ symmetry. Both the origin and $\bm{r}_{BB}$ are points with $C_3$ symmetry. For the SE-even structures there are
            \begin{align}
            N = \frac{4}{3} \frac{m^2+m^2+mn}{\text{gcd}(m,n)^2}
            \end{align}
            atoms per unit cell.
            
            Complementarily, the SE-odd structures obey
            \begin{align}
            \frac{n-m}{\text{gcd}(m,n)}\ \text{mod}\ 3 = 1 \text{ or } 2,
            \end{align}
            and are likewise rotated about a point where an atom in the $A$ sublattice of the second layer lies above an atom of the $A$ sublattice in the first layer. This point has $C_3$ symmetry and is the point of greatest symmetry in the lattice. The structures $(m,n)$ with $(n-m)/\text{gcd}(m,n)\text{ mod }3=1$ and $(m',n')=(n,m)$ with $(n'-m')/\text{gcd}(m,n)\text{ mod }3=2$ are related by an in-plane reflection.
            SE-odd structures are related to commensuration partner SE-even structures by a change in twist angle $\theta_\text{odd}=\pi/3-\theta_\text{even}$ \cite{mele2010commensuration}, an in-plane translation \cite{talkington2023electric}, or in terms of $m$ and $n$ \cite{mele2012interlayer} 
            \begin{align}
            \begin{pmatrix}
            m\\n
            \end{pmatrix}_\text{odd} = \frac{1}{3}\begin{pmatrix} -1 & 1\\ 3-1 & 1\end{pmatrix} \begin{pmatrix}
            m\\n
            \end{pmatrix}_\text{even}
            \end{align}
            For SE-odd structures there are
            \begin{align}
            N = 4 \frac{m^2+m^2+mn}{\text{gcd}(m,n)^2}
            \end{align}
            atoms per unit cell.

        \subsection{Real Space Translation and Momentum Space Structure}

            \begin{figure}
                \centering
                \includegraphics[width=\linewidth]{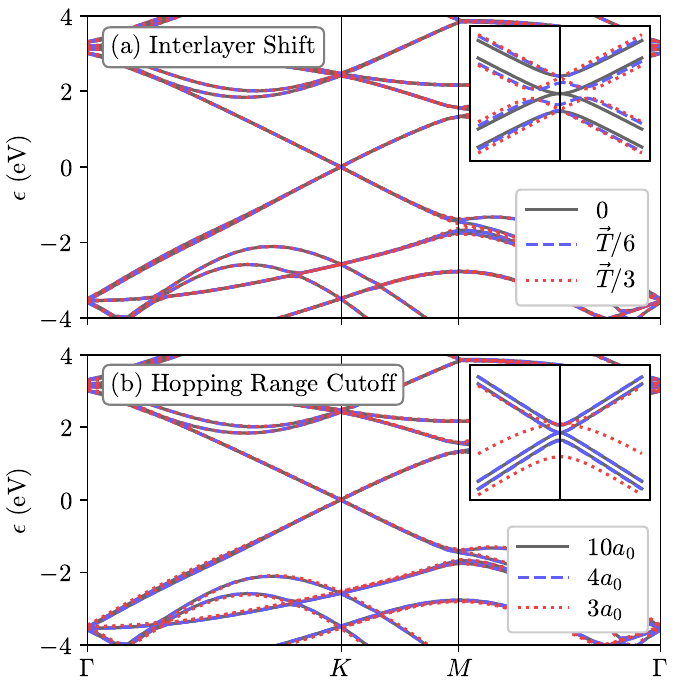}
                \caption{\textbf{(a)} Interlayer shift determines the low energy band structure, but the high energy band structure is invariant under shifts; visualized for the $N=28$ structure where $\vec{T}=(\bm{a}^M_1+\bm{a}^M_2)/7$. \textbf{(b)} While short range hoppings are sufficient to describe high energy degrees of freedom, for the $N=28$ structures sixth nearest neighbors ($4a_0$) is necessary to capture the low energy behavior. Longer range hoppings are needed to describe structures with larger unit cells.}
                \label{fig:cutoff}
            \end{figure}

            In the superlattice heterostructure community, the prevailing view is that intralayer superlattice translations are nearly irrelevant for electronic physics \cite{bistritzer2011moire}. While this may be true at sufficiently large energies and at small energies in systems with large unit cells, this is crucially not the case for large twist angle C-TBG \cite{moon2013optical}. In particular, the low energy physics occurs exclusively on a locus of points surrounding the $K$ and $K'$ points \cite{mele2010commensuration}. This localization of states that contribute to the low energy properties in momentum space means that the long range real space structure is important for determining the low energy electronic physics. In Fig. \ref{fig:cutoff}, we illustrate the importance of interlayer translation in C-TBG at low energies by contrasting the low energy and band structures for C-TBG as a dependent on interlayer translation, and as dependent on tight-binding hopping range cutoff.

            In the SE-odd structure, the low energy band structure is characterized by a quadratic band touching, as in the special case of the Bernal bilayer. This quadratic dispersion originates from the combination of layer Dirac fermions in $l=\hbar$ angular momentum eigenstates to fermions in an $l=2\hbar$ eigenstate. This angular momentum results in a phase winding that leads to an exact cancellation of the momentum-integrated Hall conductivity at low energies as visualized in Fig. \ref{FIG:real28}(c). To obtain terahertz circular dichoism in these materials the low energy states must not be even-integer valued angular momentum eigenstates. This can be achieved through interlayer translation or the application of an electric field, as visualized in Fig. \ref{FIG:real28}(d).

            \begin{table}
            \textbf{Interlayer Coherence Scale in C-TBG}
            \begin{tabular}{c|c|c|c|c|c}
                $N$ (atoms) & $(m,n)$ & Twist Angle & $2V_0$ (DFT) & $2V_0$ (TB) & $\varphi$ (TB) \\\hline
                4 & (1,1) & $30-30^\circ$ & --- & 675.6 & 0$^\circ$ \\
                28 & (1,4) & $30+8.21^\circ$ & 7.0 & 5.737 & 70.14$^\circ$ \\
                52 & (2,5) & $30-2.20^\circ$  & 0.2 & 0.033 & 109.67$^\circ$ \\
                76 & (1,7) & $30+16.83^\circ$ & 0.4 & 0.199 & 91.68$^\circ$ \\
                124 & (4,7) & $30-12.10^\circ$ & 0.1 & 0.004 & 38.26$^\circ$ \\
                148 & (1,10) & $30+20.57^\circ$ & 0.0 & 0.008 & 99.45$^\circ$\\
                172 & (5,8) & $30-14.82^\circ$ & --- & 3.4e-5 & 119.12$^\circ$ \\
                196 & (2,11) & $30+13.57^\circ$ & --- & 5.3e-7 & 135.64$^\circ$ \\
                244 & (1,13) & $30+22.66^\circ$ & --- & 3.3e-4 & 103.85$^\circ$\\
                268 & (5,11) & $30-5.57^\circ$ & --- & 4.0e-7 & 103.63$^\circ$\\
                292 & (7,10) & $30-18.36^\circ$ & --- & 6.3e-6 & 25.11$^\circ$\\
                316 & (4,13) & $30+3.99^\circ$ & --- & 7.9e-9 & 62.83$^\circ$\\
                364 & (8,11) & $30-19.58^\circ$ & --- & 7.6e-8 & 118.30$^\circ$\\
                388\footnote{The $N=388$ case is on the edge of 64-bit numeric precision.} & (5,14) & $30+0.59^\circ$ & --- & 1e-12 & 107.96$^\circ$ \\
            \end{tabular}
            \caption{Energy separation, $2V_0$, of the Dirac points in SE-even structures due to interlayer scattering measured in meV as determined using density functional theory (DFT) and tight-binding methods. Structures with $N\leq 388$ atoms per unit cell are considered. The interlayer coherence scale $V_0$ decays rapidly with unit cell size and the interlayer coherence scale is only in the terahertz region (1 meV = 0.242 THz) for the 28 atom unit cell. DFT calculations are from Ref. [\onlinecite{shallcross2008quantum}]. The phase of the scattering $\varphi$ is also determined through fitting the band gap minimum to a symmetry based model.}\label{TABLE:coherence}
            \end{table}

            \begin{figure}
                \centering
                \includegraphics[width=\linewidth]{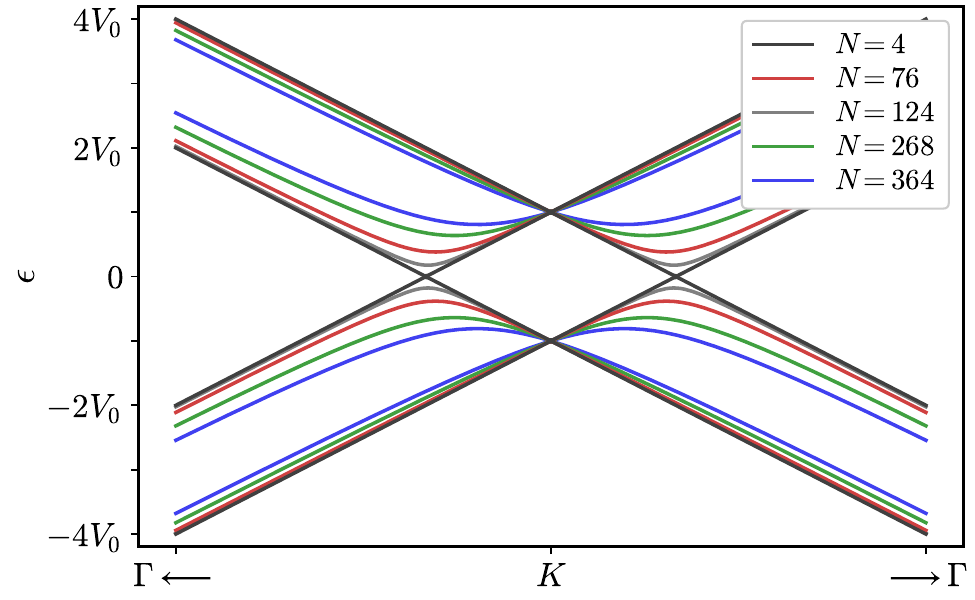}
                \caption{Representative low energy band structures of SE-even structures around the $K$ point. Energies are expressed in terms of the interlayer coherence $V_0$ so that the structures overlap and their universal Dirac behavior at large energies is manifest. The structures are \textit{not} all the same since the twist angle and the phase $\varphi$ changes the band gap $2V_0\sin((\varphi-\theta)/2)$ and low energy dispersion. The $N=4$ structure (AA bilayer, $\theta=0$) is gapless while the largest gap for the structures with less than 400 atoms per unit cell is $1.617 V_0$ for the $N=364$ structure.} \label{FIG:even-bands}
            \end{figure}
            
        \subsection{Tight-Binding Model and Interlayer Coherence}

            Since these structures are commensurate and hence periodic in real-space they permit a momentum space description. While at small incommensurate angles it is possible to model the low energy physics twisted bilayer graphene as that of two decoupled Dirac cones, this description breaks down since in the SE-even structures the $K$ valleys of the two layers are separated by a reciprocal lattice vector, while in the SE-odd structures the $K$ valley of one layer and the $K'$ valley of the other layer are separated by a reciprocal lattice vector \cite{mele2010commensuration,mele2012interlayer}.

            These interlayer scattering processes have an amplitude $V_0$ and modify the low energy band structure over a scale proportional to this interlayer coherence scale $V_0$. Now it has been found that for small commensurate unit cells $V_0$ decays rapidly with unit cell size \cite{shallcross2008quantum}. Here we use a tight binding model to determine $V_0$ for all commensurate structures with less than 400 atoms. The real space structures of the eight smallest SE-even bilayers are plotted in Fig. \ref{FIG:realeven}; the SE-odd structures also form at the same twist angle but with an interlayer translation.
            
            We use the two-center Slater-Koster type model of $p_z$ orbitals on carbon atoms with hopping $t(\bm{r}) = V_{pp\pi}(\bm{r}) + V_{pp\sigma}(\bm{r})$ developed and used in Refs. [\onlinecite{trambly2010localization}] and [\onlinecite{moon2013optical}], where the hopping terms are
            \begin{align}
                V_{pp\pi}(\bm{r}) &= t_{pp\pi} e^{-(|\bm{r}|-a_0)/\delta} \left(1-\left(\frac{\bm{r}\cdot \bm{e}_z}{|\bm{r}|}\right)^2\right)\\
                V_{pp\sigma}(\bm{r}) &= t_{pp\sigma} e^{-(|\bm{r}|-d)/\delta}\left(\frac{\bm{r}\cdot \bm{e}_z}{|\bm{r}|}\right)^2
            \end{align}
            for hopping magnitudes $t_{pp\pi} = 2.7$ eV and $t_{pp\sigma} = -0.48$ eV, unit vector perpendicular to the bilayer $\bm{e}_z = (0,0,1)$, intra-layer atomic spacing $a_0 = a/\sqrt{3} = 0.142$ nm, inter-layer spacing $d = 0.335$ nm, and decay length $\delta = 0.184 a = 0.0453$ nm. We provide the atomic positions for SE-even structures in the Supplementary Information \footnote{Supplementary Information to be included in published version}.

            As expected, we find that there are only four bands at low energies corresponding to the layer and sublattice degrees of freedom. We plot the bands for representative SE-even structures in Fig. \ref{FIG:even-bands}. We determine the interlayer coherence by the band gap at the $K$ point, where the band gap is $2V_0$. Additionally there is a symmetry allowed phase shift $\varphi$ for intravalley interlayer scattering (in SE-even structures), which we determine through the band gap minimum which is given by $2V_0\sin[(\varphi-\theta)/2]$ where $\theta$ is the twist angle \cite{talkington2023electric}. We tabulate the results in Table \ref{TABLE:coherence}.

        \begin{figure*}
            \includegraphics[width=\linewidth]{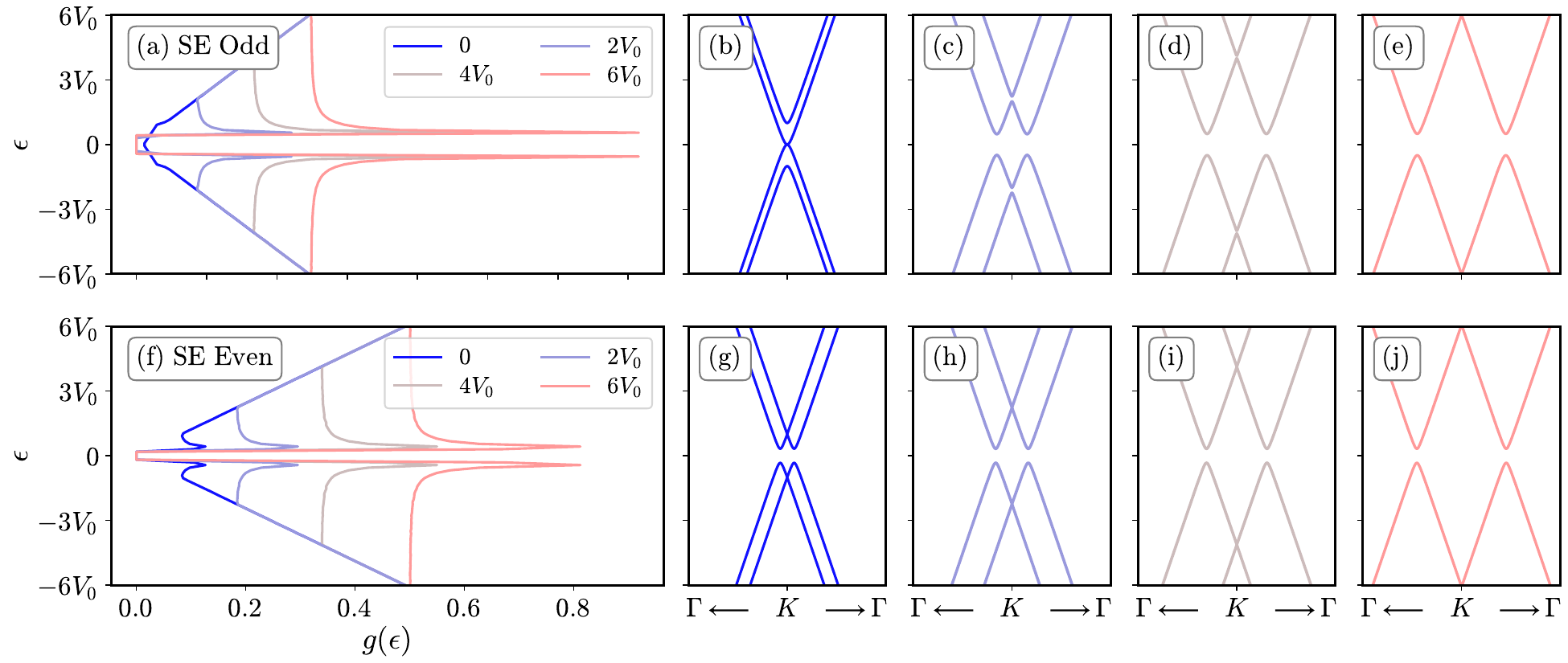}
            \caption{Densities of states and band structures for SE-odd and SE-even structures with perpendicular electric field strengths from $0$ to $12V_0/ed\approx 0.1028$ V/nm. \textbf{Top Row:} \textbf{(a)} The SE-odd structure is gapless in the absence of an electric field, but develops a gap that rapidly saturates as $V_0\sqrt{x^2/(1+x^2)}$ where $x=4\mathcal{E}/V_0^2$ and to van Hove singularities develop at the band edges. \textbf{(b)} SE-odd structure in the absence of an electric field is gapless with a quadratic band touching similar to the Bernal bilayer. \textbf{(c-e)} In the presence of an electric field, SE-odd band structures are gapped and exhibit an avoided crossing above and below the Fermi energy corresponding to the field decoupling the layers at the $K$ point. \textbf{Bottom Row:} \textbf{(f)} The SE-even structures are always gapped with magnitude $2V_0\sin((\varphi-\theta)/2)$ for twist angle $\theta$. In the presence of an electric field, the Dirac points move away from the Fermi energy and the the band gap minima move away from the $K$ point linearly in field strength leading to an enhancement of the density of states at low energies and a delayed onset of universal Dirac cone behavior. \textbf{(g-e)} SE-even structures feature a constant gap and the Dirac cone separation scales linearly with the electric field.}\label{fig:oddevenDOS}
        \end{figure*}

        \subsection{Low-Energy Continuum Model}

            Now we have seen that the low energy behavior of C-TBG consists of four bands near the Fermi energy corresponding to the layer and sublattice degrees of freedom. These four bands are the layer Dirac cones hybridized by interlayer scattering. If the systems' real-space structure has points with $C_6$ symmetry it is SE-even while if it has points with $C_3$ symmetry it is SE-odd. In general for an arbitrary interlayer shift there will be no points with $C_6$ or $C_3$ symmetry, but the behavior in between these sublattice symmetric structures interpolates smoothly.
            
            In a previous work one of us derived a symmetry based model for these extremal cases of SE-odd and SE-even structures \cite{mele2010commensuration}. The model consisted of two layer Dirac fermions with a relative interlayer twist and symmetry-allowed interlayer tunneling terms. The SE-odd structures permit interlayer intravalley scattering while the SE-even structures permit interlayer intervalley scattering with a phase shift.
            Explicitly the Bloch Hamiltonians are for a momentum $\bm{k}$ away from the $K$ point
            \begingroup
            \setlength\arraycolsep{0.3pt}
            \begin{align}
                H_{\bm{k}}^\text{odd} = \begin{pmatrix}
                \mathcal{E}+\mu & k_x - ik_y & V_0 & 0\\
                \!k_x + ik_y & \mathcal{E}+\mu & 0 & 0\\
                V_0 & 0 & -\mathcal{E}+\mu & \!\!-e^{i\theta}(k_x+ik_y)\!\\
                0&0&\!\!-e^{-i\theta}(k_x-ik_y)\!\!&-\mathcal{E}+\mu
                \end{pmatrix}
            \end{align}
            and
            \begin{align}
                H_{\bm{k}}^\text{even} = \begin{pmatrix}
                \mathcal{E}+\mu & k_x - ik_y & V_0 e^{i\varphi/2} & 0\\
                \!k_x + ik_y & \mathcal{E}+\mu & 0 & V_0 e^{-i\varphi/2}\\
                V_0 e^{-i\varphi/2} & 0 & -\mathcal{E}+\mu & e^{-i\theta}(k_x-ik_y)\!\\
                0&V_0 e^{i\varphi/2}\,\,&e^{i\theta}(k_x+ik_y)\!\!&-\mathcal{E}+\mu
                \end{pmatrix}
            \end{align}
            \endgroup 
            where the $\mathcal{E}$ term corresponds to a perpendicular electric field and the $\mu$ term is the chemical potential. In what follows we shall use these symmetry based models to compute the optical response properties of the SE-odd and SE-even structures since for the low energies we are interested in the difference from these and the tight-binding models is negligible.

        \subsection{Properties of SE-odd and SE-even Structures}
            
            The SE-odd and SE-even structures return to the behavior of two decoupled layer Dirac cones outside of the energy window given by $V_0$ and $\mathcal{E}$. This universal behavior is characterized by a linear dispersion and a linearly increasing density of states (DOS) as illustrated in Fig. \ref{fig:oddevenDOS}. Below this universal regime the structures are richer.
            
            The SE-odd structure is gapless but becomes gapped when a perpendicular electric field is applied and exhibits van Hove singularities at the band edge. Above the van Hove singularity the DOS flattens and until the system enters the universal Dirac regime above the energy of the band gap at $K$.

            Meanwhile, the SE-odd structure is gapped, but the band gap at the $K$ point increases when a perpendicular electric field is applied. The system exhibits van Hove singularities at the band edge. Like in the SE-odd case, above the van Hove singularity the DOS flattens and until the system enters the universal Dirac regime above the energy of the band gap at $K$.

        \begin{figure*}
                \includegraphics[width=\linewidth]{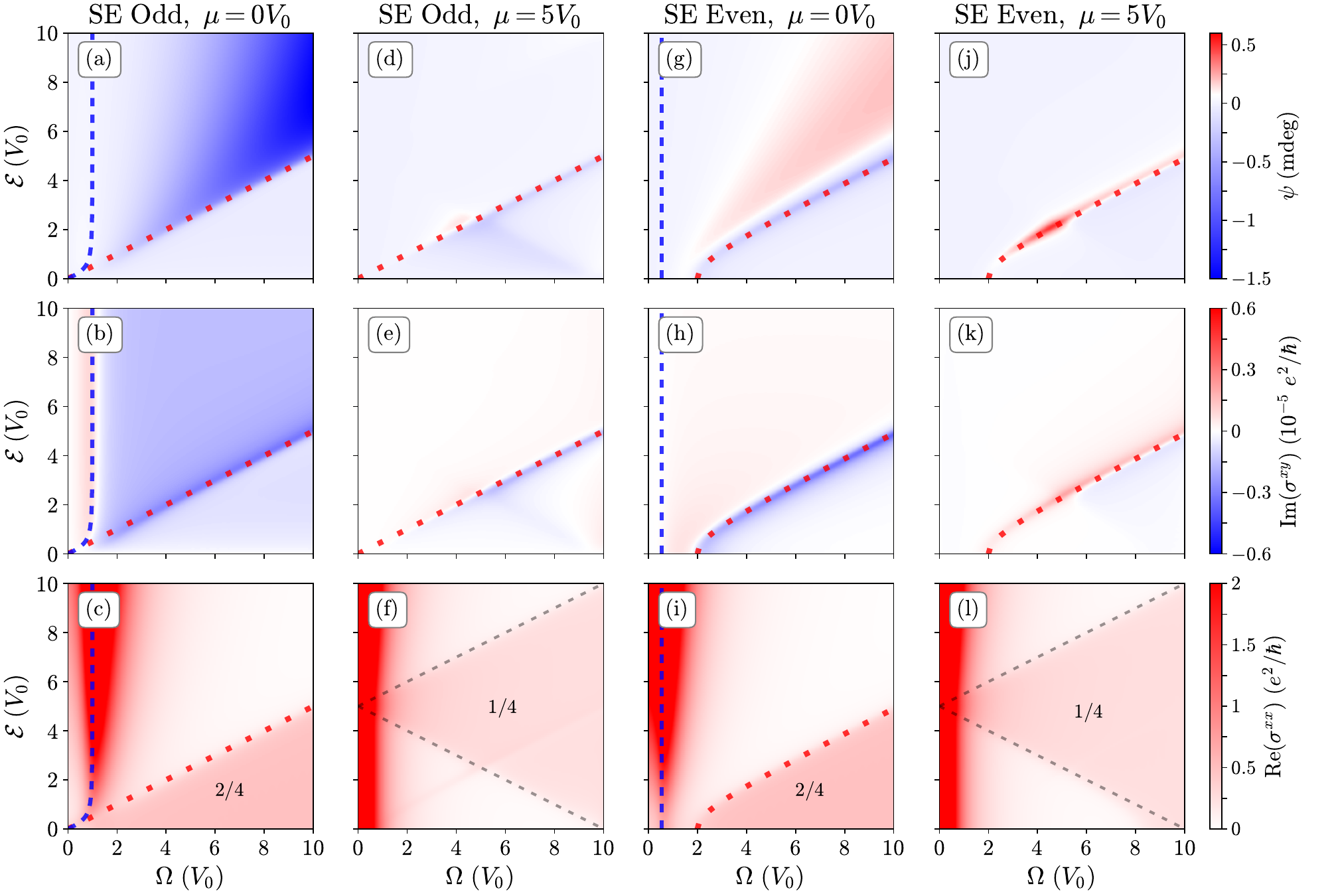}
                \caption{Ellipticity $\psi$ and optical conductivity $\sigma$ in SE-odd (21.79$^\circ$) and SE-even (38.21$^\circ$) C-TBG as a function of frequency and perpendicular electric field. The undoped systems exhibit sizeable circular dichroism at large electric field and a corresponding frequency as seen in the dark blue/red regions in panels (a) and (g). For $2V_0=5.737$ meV, the frequencies range from 0 to 6.9 THz and the applied fields range from 0 to 0.17 V/nm. Up to a change in the scattering phase $\varphi$, the results can be applied to other commensurate structures by changing $V_0$. The dashed blue lines are the band gap minimum, the dashed red lines are the band gap at $K$, and the dashed gray lines are $\mu\pm \Omega/2$. \textbf{(a)} Ellipticity is largest in the region before universal Dirac behavior at large fields and grows as $\Omega^2$ to a peak that scales as $\mathcal{E}$, \textbf{(b)} the imaginary part of the Hall conductivity is positive below the band gap minimum, negative above it but below the band gap at $K$, and tends to zero in the universal Dirac regime, \textbf{(c)} the real part of the longitudinal conductivity peaks at the band gap minimum and exhibits a power law divergence \cite{talkington2023electric} before decaying as $\Omega^{-2}$ when the transitions are between regions with flat densities of states, and finally exhibiting twice the universal Dirac conductivity for the bilayer. \textbf{(d)} Upon doping, the hybridized states at low energy are filled and do not contribute to circular dichroism, \textbf{(e)} the only non-vanishing part of $\text{Im}(\sigma^{xy})$ is at the band gap at the $K$ point which only probes the low energy behavior, \textbf{(f)} the doped systems exhibit a Drude peak at low frequency and a sub-universal plateau at $e^2/4\hbar$ corresponding to photons probing only one of the Dirac cones; at higher frequencies transitions probe both Dirac cones and the conductivity becomes $2e^2/4\hbar$. \textbf{(g)} As in the SE-odd case the ellipticity is largest in the region before the universal Dirac regime, and the ellipticity grows as $\Omega^2$ to a peak that scales with $\mathcal{E}$, \textbf{(h)} $\text{Im}(\sigma^{xy})$ exhibits the same features as in the SE-odd structure, but with flipped sign, and an additional crossover in sign as the system approaches universal Dirac behavior, \textbf{(i)} $\text{Re}(\sigma^{xy})$ exhibits the same features as in the SE-odd structure once the different band gap is taken into account. \textbf{(j-l)} Upon doping, the behavior of the SE-even structure is similar to that of the SE-odd structure once one accounts for the shifted band gap, where only the transitions probing the gap at $K$ contribute to the ellipticity.}\label{FIG:ellipticity}
            \end{figure*}

            \begin{figure}
                \centering
                \includegraphics[width=\linewidth]{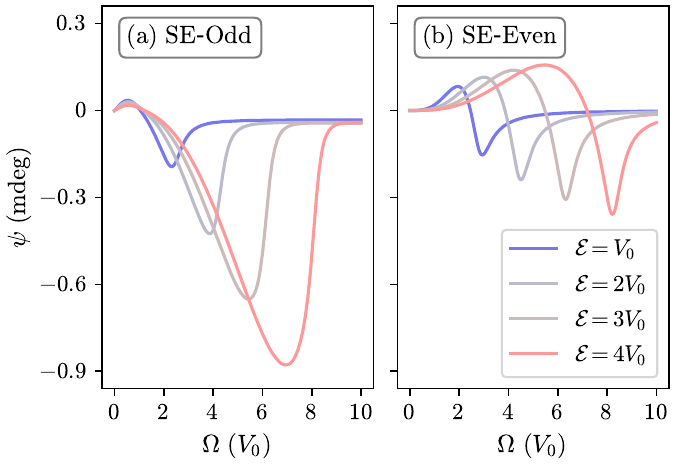}
                \caption{Ellipticity $\psi$ in undoped, SE-odd \textbf{(a)}, and SE-even \textbf{(b)}, structures as dependent on perpendicular applied field. Note that for field strengths above the interlayer coherence scale $V_0$ the ellipticity scales roughly linearly in field.}
                \label{FIG:cd-slices}
            \end{figure}

    \section{Linear Response Formalism}\label{sec:response}

        We use the Kubo formula for linear response to determine the response of C-TBG to linearly and circularly polarized light. The optical conductivity is \cite{mahan2000many}
        \begin{align}
                \sigma^{\mu\nu}(\Omega) = \frac{i}{\Omega} \sum_{s,s'} \int_{BZ} \frac{d^2\bm{k}}{(2\pi)^2} \frac{(f(\epsilon_s)-f(\epsilon_{s'}))J^\mu_{ss'}J^\nu_{s's}}{\Omega - (\epsilon_{s'}-\epsilon_s) + i\eta}.
        \end{align}
        $\Omega$ is the photon frequency, $f(\epsilon)=1/(e^{\beta(\epsilon+\mu)}+1)$, $J^\mu_{ss'}=\langle u_s|J^\mu|u_{s'}\rangle$, and $\eta$ is a phenomenological broadening term. The key modification to this formalism from that for purely two dimensional systems is to account for the thickness of the bilayer and the resulting phase shift of the light in the current operators.

        \subsection{Current Operators}

            We consider a current operator that is composed of currents in the top and bottom layers with a phase shift for the separation of the layers $d$ as first presented by Morrell et al in Ref. [\onlinecite{morell2017twisting}]
            \begin{align}
                J^\mu = J_T^\mu e^{i\frac{\Omega}{c}d} + J_B^\mu,
            \end{align}
            where the current operators in the layers are $J_{T(B)}^\mu = P_{T(B)} (\partial H_{\bm{k}}/\partial k_\mu)$ where $P_{T(B)}$ are the projectors onto the sites in the top layer $T$ and bottom layer $B$ respectively.
            We consider the response to first order, and $\Omega d<\!\!<c$ so we expand the exponential to linear order
            \begin{align}
                J^\mu = J_T^\mu(1+i\tfrac{\Omega}{c}d) + J_B^\mu.
            \end{align}
            For the longitudinal components of $\sigma$, the response to zeroth order is non-vanishing, while the first order response vanishes due to time-reversal symmetry (TRS) \cite{morell2017twisting}, so the matrix elements are
            \begin{align}
                J^\mu_{ss'}J^\mu_{s's} = \sum_{l,l'\in\{T,B\}} \langle u_s|J^\mu_l|u_{s'}\rangle\langle u_{s'}|J^\mu_{l'}|u_s\rangle,
            \end{align}
            which for $J_\text{lin}^\mu = \partial H_{\bm{k}}/\partial k_\mu$ gives the typical expression
            \begin{align}
                J^\mu_{ss'}J^\mu_{s's} = |\langle u_s|J^\mu_\text{lin}|u_{s'}\rangle|^2.
            \end{align}
            
            Now, the transverse components vanish to zeroth order since C-TBG is non-magnetic \cite{morell2017twisting} so we consider the first order contributions
            \begin{align}
                J^\mu_{ss'}J^\nu_{s's} &= i\tfrac{\Omega}{c}d \langle u_s| J_T^\mu | u_{s'}\rangle\langle u_{s'}| J_T^\nu|u_s\rangle\nonumber\\ &+ i\tfrac{\Omega}{c}d \langle u_s| J_T^\mu | u_{s'}\rangle\langle u_{s'}| J_B^\nu|u_s\rangle\nonumber\\ & + i\tfrac{\Omega}{c}d \langle u_s| J_B^\mu | u_{s'}\rangle\langle u_{s'}|J_T^\nu|u_s\rangle.
            \end{align}
            When $\mu\neq \nu$ we have $\sum_{s,s'}\langle u_s| J_{T}^\mu | u_{s'}\rangle\langle u_{s'}|J_{T}^\nu |u_s\rangle = 0$ since the monolayer has no thickness. Swapping indices, accounting for the Fermi functions, and that in the presence of TRS $\sum_{ss'} \langle u_s|J_T^\nu|u_{s'}\rangle\langle u_{s'}|J_B^\mu|u_s\rangle = -\sum_{ss'} \langle u_s|J_T^\mu|u_{s'}\rangle\langle u_{s'}|J_B^\nu|u_s\rangle$ for $\mu\neq\nu$, we have \cite{morell2017twisting,stauber2018chiral}
            \begin{align}
                J^\mu_{ss'}J^\nu_{s's} &= 2i\tfrac{\Omega}{c}d \langle u_s| J_T^\mu | u_{s'}\rangle\langle u_{s'}| J_B^\nu |u_s\rangle,
            \end{align}
            where summing this expression over will give us the same linear order Hall response as for the full current operators.

    \section{Circular Dichroism and Optical Conductivity}\label{sec:cd}

            An experimentally accessible measure of circular dichroism and natural optical activity is ellipticity which measures the conversion of linearly polarized light to elliptically polarized light \cite{morell2017twisting,stauber2018chiral,addison2019twist}. In terms of the optical conductivity the ellipticity is
            \begin{align}
            \psi = \frac{\text{Im}(\sigma^{xy})}{2\,\text{Re}(\sigma^{xx})}
            \end{align}
            which is large when $\text{Im}(\sigma^{xy})$ is large and $\text{Re}(\sigma^{xx})$ is small. For an analysis of the ellipticity that takes into account the finite thickness of the bilayer, see Appendix \ref{appendix:ellipticity}.
            
            In the pre-universal Dirac regime, the application of an electric field results in a flat density of states for an energy range about the Fermi energy which leads to $\text{Re}(\sigma^{xx})$ falls off as $\Omega^{-2}$. Combined with $\text{Im}(\sigma^{xy})$ that is constant in $\Omega$ results in an ellipticity that grows as $\Omega^2$ in the presence of a field. The application of larger fields extend the range of this pre-universal regime and the peak magnitude scales linearly in $\mathcal{E}$.
            At visible photon energies the application of an electric field can also be used to enhance and tune the circular dichroism, although the field dependence is complicated by the number of bands available to resonant transitions.

            Now we have expressions that we can integrate to obtain the optical conductivity. The expression for the Hall conductivity includes the effect of the phase difference of light in the two layers which is the origin of the circular dichroism in these materials.

            We numerically integrate the continuum models using a discretized momentum space mesh for the SE-odd and SE-even models, both with and without doping as a function of photon frequency and perpendicular electric field. We consider the response at a temperature of $T=4$ K and use a broadening of $\eta=1$ meV. We plot the results in Fig. \ref{FIG:ellipticity} where the middle row and the bottom row display the imaginary part of the Hall and real part of the longitudinal conductivity respectively.

            In the undoped systems the behavior is largely set by the band gap minimum where there is a power law divergence, as reported in our previous work Ref. \cite{talkington2023electric}, and the band gap at the $K$ point which dictates the onset of universal Dirac behavior. Between these two gaps $\text{Re}(\sigma^{xx})$ falls off as $\Omega^{-2}$ while $\text{Im}(\sigma^{xy})$ is constant in $\Omega$ leading to an ellipticity that grows as $\Omega^2$ to a peak whose magnitude scales as $\mathcal{E}$. In the universal regime the $\text{Im}(\sigma^{xy})$ asymptotes to zero while $\text{Re}(\sigma^{xx})$ saturates to $2e^2/4\hbar$: twice the optical conductivity of the monolayer.
            In the doped systems the non-universal behavior is hidden in the Fermi sea and the ellipticity is suppressed and only emerges at the Fermi surface in the form of transitions at the $K$ point. The ellipticity at low frequencies is suppressed by the Drude peak. At finite but small frequencies only one Dirac cone is probed so the real part of the optical conductivity assumes the value of the monolayer. At higher frequencies the universal optical conductivity of the bilayer is recovered.

            So we find that the ellipticity is large in the THz region in C-TBG when the doping allows the non-Dirac cone structure to be probed and this ellipticity is enhanced through the application of an electric field.

    \section{Conclusion}\label{section:...}

        Circular dichoism in the absence of magnetism is suppressed as $\Omega$, so a priori it seems that terahertz circular dichroism due to natural optical activity should be vanishingly small. We show that in contrast to this picture, the coherent superposition of contributions to circular dichroism in C-TBG leads to terahertz a circular dichroism comparable to the circular dichoism of C-TBG at photon energies two orders of magnitude larger \cite{kim2016chiral,addison2019twist}. At low energies C-TBG has valley symmetry as the valleys are separated by a reciprocal lattice vector. This symmetry ensures that the contributions to the circular dichroism add instead of cancel. Additionally the magnitude of the matrix elements contributing to this response are large as a result of interlayer hybridization over a range set by an interlayer coherence scale. We determined this interlayer coherence scale for all C-TBG structures with less than 400 atoms in the unit cell and find that the 28 atom unit cell structures are the most conducive to the presentation of terahertz circular dichoism.
        
        The circular dichroic response can be enhanced through the application of an electric field, and is suppressed with doping. The strength of applied electric fields we studied here are comparable to those applied in real devices \cite{ma2022intelligent}, and the frequencies studied here are accessible to contemporary photon sources.
        Circularly polarized terahertz light may be useful for identifying chiral chemical molecules with low energy excitations such as vibrational modes.
        We present commensurate twisted bilayer graphene as a tunable platform to achieve large terahertz circular dichroisms in the absence of magnetism.
    
    \begin{acknowledgements}
        S.T. acknowledges support from the NSF under Grant No. DGE-1845298. E.J.M. acknowledges support from the DOE under Grant No. DE-FG02-84ER45118.
    \end{acknowledgements}

    \appendix

    \section{Effect of Finite Bilayer Thickness}\label{appendix:ellipticity}

\begin{figure}
        \centering
        \includegraphics[width=0.85\linewidth]{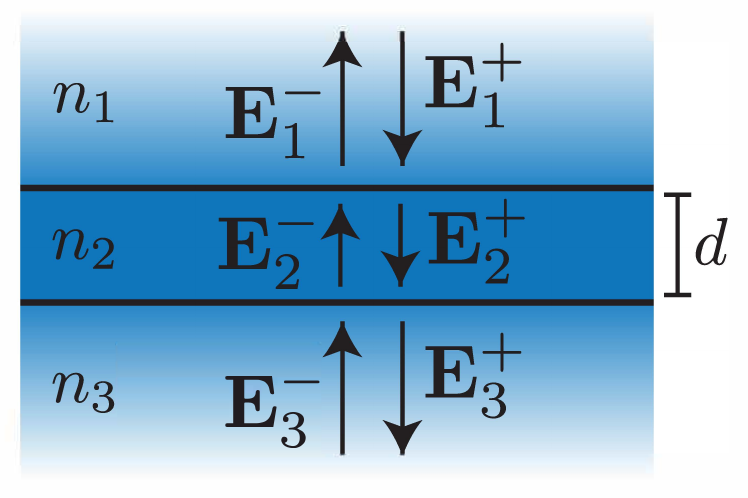}
        \caption{Geometry considered in Appendix \ref{appendix:ellipticity}. Finite thickness of the bilayer is treated by using a transfer matrix method.}
        \label{fig:bilayer-finite}
    \end{figure}

    Now, the analysis presented in the main body of the paper is based on an approximate solution to Maxwell's equations \cite{morell2017twisting,stauber2018chiral,addison2019twist}. A more precise formulation incorporates the boundary effects on the bilayers \cite{kuzmenko2009determination,szechenyi2016transfer,li2018two,ho2023optical}. We present the salient features from Ref. \cite{ho2023optical}, and work in units of conductivity where $\sigma_0 = e^2/4\hbar$. The natural quantity to investigate are the electric field amplitudes on the two sides of the bilayer
    \begin{align}
    \begin{pmatrix}
        \bm{E}_3^+(z_2)\\
        \bm{E}_3^-(z_2)
    \end{pmatrix}
    = M_{31}
    \begin{pmatrix}
        \bm{E}_1^+(z_1)\\
        \bm{E}_1^-(z_1)
    \end{pmatrix},
    \end{align}
    where $\bm{E}^\pm$ are propagating from (towards) the light source, and 
    \begin{align}
    M_{31} = M_{32}^BM_{32}^AM_2M_{21}^BM_{21}^A
    \end{align}
    is the transfer matrix that takes fields from one side of the bilayer to the other. See Fig. \ref{fig:bilayer-finite} for an illustration of the bilayer and its three regions.
    Now the transfer matrices for the electric field at normal incidence are given by \cite{ho2023optical}
    \begin{widetext}
        \begin{align}
        M_{32}^B = \begin{pmatrix}
        \tau^0 & \tau^0\\
        - \frac{n_3}{2\alpha}\sigma_0\tau^0 - \sigma^{xx}_{(2,2)}\tau^0 & \frac{n_3}{2\alpha}\sigma_0\tau^0 - \sigma^{xx}_{(2,2)}\tau^0
        \end{pmatrix}^{-1}
        \end{align}
        \begin{align}
        M_{32}^A =
        \begin{pmatrix}
        \tau^0 & \tau^0\\
        - \frac{n_2}{2\alpha}\sigma_0\tau^0 + (\sigma^{xx}_{(1,2)}\tau^0 -i \sigma^{xy}_{(1,2)}\tau^2)^\dagger e^{-in_2\Omega d/c} & \frac{n_2}{2\alpha}\sigma_0 \tau^0 + (\sigma^{xx}_{(1,2)}\tau^0 -i \sigma^{xy}_{(1,2)}\tau^2)^\dagger e^{in_2\Omega d/c}
        \end{pmatrix}
        \end{align}
        \begin{align}
        M_2 = \begin{pmatrix}
            e^{in_2\Omega d/c} \tau^0 & 0\\
            0 & e^{-in_2\Omega d/c} \tau^0
        \end{pmatrix}
        \end{align}
        \begin{align}
        M_{21}^B =
        \begin{pmatrix}
        \tau^0 & \tau^0\\
        - \frac{n_2}{2\alpha}\sigma_0\tau^0 - (\sigma^{xx}_{(1,2)}\tau^0 -i \sigma^{xy}_{(1,2)}\tau^2)e^{in_2\Omega d/c} & \frac{n_2}{2\alpha}\sigma_0 \tau^0 - (\sigma^{xx}_{(1,2)}\tau^0 -i \sigma^{xy}_{(1,2)}\tau^2) e^{-in_2\Omega d/c}
        \end{pmatrix}^{-1}
        \end{align}
        \begin{align}
        M_{21}^A = \begin{pmatrix}
        \tau^0 & \tau^0\\
        - \frac{n_1}{2\alpha}\sigma_0\tau^0 + \sigma^{xx}_{(1,1)}\tau^0 & \frac{n_1}{2\alpha}\sigma_0\tau^0 + \sigma^{xx}_{(1,1)}\tau^0
        \end{pmatrix},
        \end{align}
    \end{widetext}
    where $\tau^0$, $\tau^2$ are Pauli matrices, $\alpha\approx1/137$ is the fine-structure constant, $n_1=n_2=n_3=1$ are the indices of refraction, and $\sigma^{\mu\nu}_{(l,l')}$ is the conductivity generated by $J^\mu_l$ and $J^\nu_{l'}$. Note the addition of a longitudinal ``drag" term $\sigma^{xx}_{(1,2)}$ which vanishes in the approximations of the main manuscript.

    \begin{figure*}
        \centering
        \includegraphics[width=\linewidth]{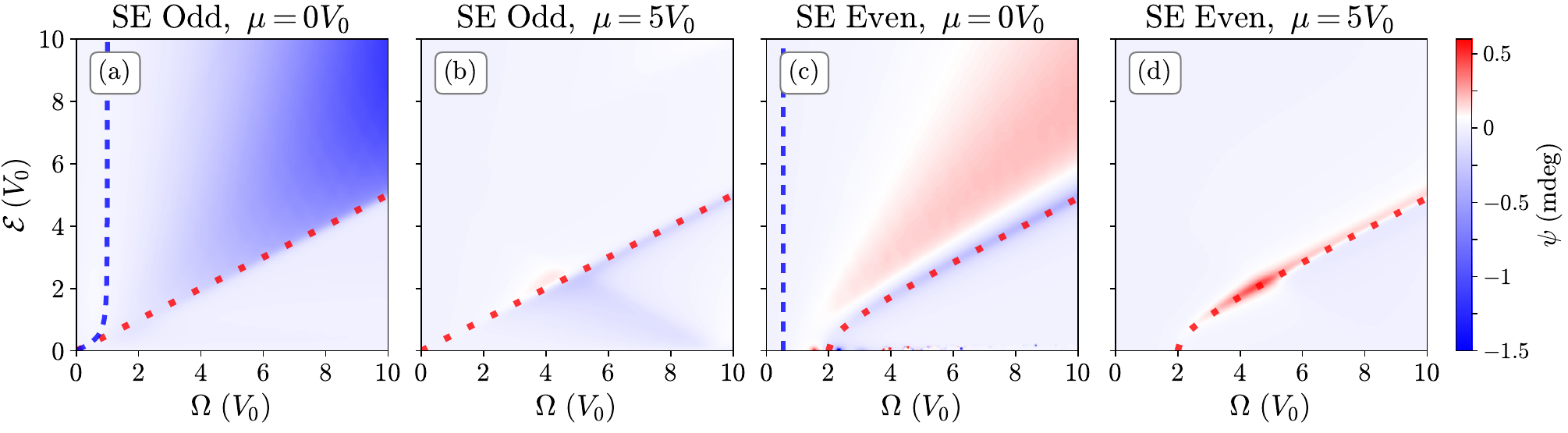}
        \caption{Ellipticity of the commensurate bilayers as dependent on photon energy $\Omega$ and perpendicular electric field strength $\mathcal{E}$, accounting for finite thickness of the bilayer. Note the strong quantitative agreement with the results when the bilayer thickness is taken to be small, c.f. Fig. \ref{FIG:ellipticity}.}
        \label{fig:refined-ellipticity}
    \end{figure*}

    From these expressions we can obtain the reflection and transmission coefficients
    \begin{align}
    r &= -(M_{31}^{22})^{-1}M_{31}^{21}\\
    t &= -(M_{31}^{-1})^{22},
    \end{align}
    where $1,2$ superscripts refer to the corresponding $2\times 2$ blocks of $M_{31}$.
    The reflectance and transmittance are
    \begin{align}
    R &= \frac{|r\bm{E}_1^+|^2}{|\bm{E}_1^+|^2}\\
    T &= \left|\frac{n_3}{n_1}\right| \frac{|t\bm{E}_1^+|^2}{|\bm{E}_1^+|^2},
    \end{align}
    and the absorbance is
    \begin{align}
    A = 1 - (R+T).
    \end{align}
    
    We consider these quantities for $\bm{E}_L=(1,i)/\sqrt{2}$ and $\bm{E}_R=(1,-i)/\sqrt{2}$, where the circular dichroism is
    \begin{align}
    \text{CD} = \frac{A_L-A_R}{A_L+A_R},
    \end{align}
    and the ellipticity is
    \begin{align}
    \psi = \frac{\Omega d}{2c}\,\text{CD}.
    \end{align}

    When we calculate this ellipticity, we find that it has all of the same qualitative characteristics as exhibited in the main text Fig. \ref{FIG:ellipticity}, as visualized in Fig. \ref{fig:refined-ellipticity}, but that the magnitude of peaks are slightly rescaled due to the drag conductivity $\sigma^{xx}_{(1,2)}$.

%

\end{document}